\documentclass[prc,preprint,floatfix]{revtex4}
\usepackage{graphicx}
\usepackage{latexsym}
\usepackage{dcolumn}
\usepackage{epstopdf}

\begin{document}

\title{Open quantum systems and Random Matrix Theory}
\author{Declan Mulhall}
 \affiliation{Department of Physics/Engineering,
 University of Scranton, Scranton, Pennsylvania 18510-4642, USA.}
 \email{mulhalld2@scranton.edu}
\date{\today}
\begin{abstract}
A simple model for open quantum systems is analyzed with Random Matrix Theory. The system is coupled to the continuum in a minimal way. In this paper the effect of opening the system on the level statistics is seen. In particular the $\Delta_3(L)$ statistic, the width distribution  and the level spacing are examined as a function of the strength of this coupling. The emergence of a super-radiant transition is observed. The level spacing and $\Delta_3(L)$ statistic exhibit the signatures of missed levels or intruder levels as the super-radiant state is formed.
\end{abstract}


\maketitle

\section{\label{sec:intro}Introduction}
Random matrix theory is used to analyse chaotic quantum systems. The statistics of the discrete energies of the system can yield information on the completeness of the data or the presence of intruder states. The standard procedure is straightforward. After preparing the data by rescaling it so that the level density is unity across its whole range (a process known as ``unfolding") the  spectral statistics of the system are compared with the RMT results for the appropriate ensemble. We would like to see what the effect coupling a chaotic system to the continuum might have on the RMT statistics. The canonical example of this process is the analysis of neutron resonance data. A free neutron is incident on a target nucleus, and they combine to make an excited compound nucleus. The incident channel is but one configuration of many. The initial wave function is simple and consists solely of this component. Through a series of random collisions of the nucleons, the initial wave function ``melts" and de-excites, emitting gamma rays on the way to the ground state. The initial configuration of a free neutron and a ground state target corresponds to one of the discrete excited states of the compound nucleus. We have come to the picture of a discrete state buried in the continuum. The system is an open quantum system. The states of the compound nucleus have a width. It is the effect of the openness of the system on the level statistics that is the main question addressed in this paper.

There is a well developed method for dealing with open quantum systems. The basic structure of the model is a Hermitian Hamiltonian with coupling to the continuum  modeled by the addition of an imaginary part or doorway state, see \cite{soze88,soze92,ZeVo2003}. The energies of the original Hamiltonian acquire widths.  A common feature of these open quantum systems is the appearance of a super-radiant state. The SR state appears as the coupling to the continuum increases. There is a restructuring of the states and one special (SR) state acquires all the width.

We will make a very simple model of an open quantum system consisting of an $N\times N$ GOE matrix with an imaginary part $=\imath \kappa \sqrt{N}$ added to the diagonal. Commonly used RMT  statistics are calculated and their behavior as a function of $\kappa$ is explored. The biggest effects occur around $\kappa=1$ which is when the SR transition happens. A plot of the energies of the opened GOE matrix vs $\kappa$ consistently show the migration of a few levels to the center of the spectrum. A plot of the level density reveals a deviation from the RMT semicircle in the middle of the energy range consistent with there being more energies close to zero as $\kappa$ grows. The entropy of the states evolves also, with the SR state clearly emerging at $\kappa=1$. Next the $\Delta_3(L)$ statistic or spectral rigidity is examined. The effect of opening the system was to increase the value of the spectral rigidity. The increase was maximum at $\kappa=1$ and then decreased, but not to zero. The shape of the $\Delta_3(L)$ curves for individual opened spectra looked like those of  incomplete spectra or ones with intruder levels. In \cite{mulhall11} it was seen that the effect on $\Delta_3(L)$ of intruder levels or missed levels was the same. The problem of spurious levels is addressed in \cite{shriner07} with both $\Delta_3(L)$ and the thermodynamic internal energy. We performed a  search for missed levels on opened spectra using RMT methods, keeping in mind that a search for intruders would give the same results. The spectra are complete, but the tests suggested that there was a fraction of the levels missed. This fraction was biggest at $\kappa=1$ where it reached a value of about 3\%. The distribution of widths also undergoes a transformation at $\kappa=1$. At large values of $\kappa$ the SR state accounts for all the width and the remaining levels have widths consistent with the Porter-Thomas distribution. We note here that the full probability density of widths for symmetric complex random matrices has been derived in \cite{fyodorov99a}. The exact solution for the case when the original matrix is from the Gaussian Unitary Ensemble is treated in \cite{fyodorov96,fyodorov99b}, where the authors derive the distribution of complex eigenvalues.

In the next section we describe the system, how it is opened and the effect this has on the energies and entropies. The SR state is seen already at this stage. In Sect.~\ref{sec:dos} we look at the density of states and address issues of unfolding the spectra in anticipation of RMT analysis. This is followed in Sect.~\ref{sec:width} by an analysis of the width distribution and a look at the SR transition. The spectral rigidity is introduced in Sect.~\ref{sec:d3} and the effect of opening the system is seen. In Sect.~\ref{sec:rmt} we perform an RMT analysis on the ensemble of open spectra using three tests for missed levels and see how open systems give false positives for missed levels. We end with concluding remarks in Sect.~\ref{sec:conc}.

\section{Opening the system,  energies and entropy}
\label{sec:opening}
Open quantum systems have been treated very successfully with an effective Hamiltonian approach.  The main ingredient is the Hamiltonian of a loosely bound system connected to continuum channels via a factorizable non-Hermitian term. The details are worked out in  \cite{soze88}, \cite{soze92} and  \cite{AuZe2011}. This method provides a general framework applicable to a broad range of systems from loosely bound nuclei \cite{AuZe2011} to electron transport in nanosystems \cite{ceka09}. The approach taken here is to make the most minimal adjustment to the GOE that would mimic openness and see how the RMT results are affected. We take a GOE matrix, $H^0$,  and add an imaginary part to the diagonal elements, $H_{ii}\rightarrow H^0_{ii}-\imath\frac{\kappa}{\sqrt{N}}$, where
$\kappa$ is the the strength of the coupling. Because the matrix is random, it is sufficient to just make the replacement $H_{11}\rightarrow H^0_{11}-\imath \kappa \sqrt{N}$ and leave other matrix elements unchanged. The resulting spectrum will be a set of complex energies $\varepsilon_n=E_n+\imath \Gamma_n$

\begin{figure}
\includegraphics[width=0.8\textheight]{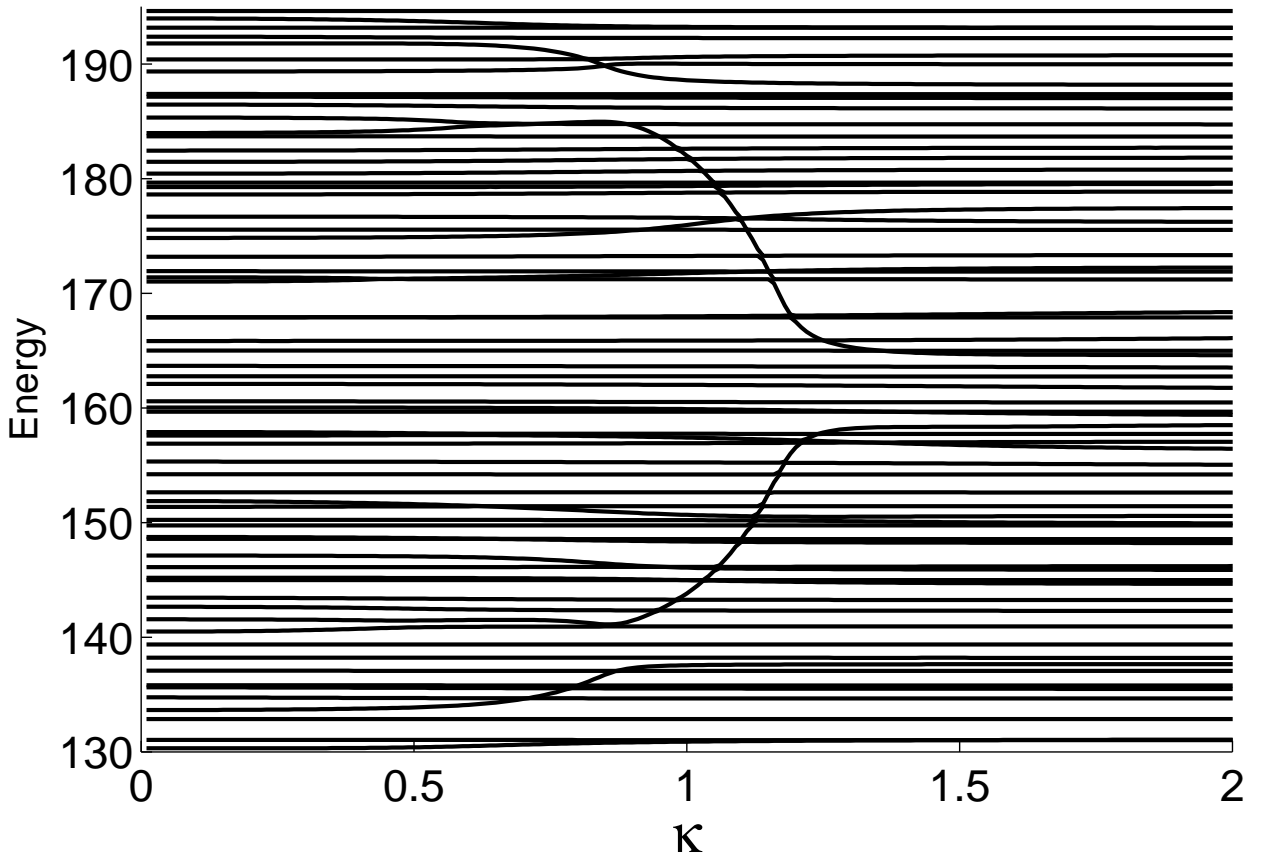}
\caption{\label{fig:evsk} The  evolution of the energy levels with
$\kappa$. We just take the 300 unfolded energies of one particular
matrix, and plot $E_n$ vs $\kappa$ for levels 130 to 195 of the
$N=300$ unfolded spectrum.  The curves, or ``trajectories" around
$\kappa=1$ vary from matrix to matrix.}
\end{figure}

\begin{figure}
\includegraphics[width=0.8\textheight]{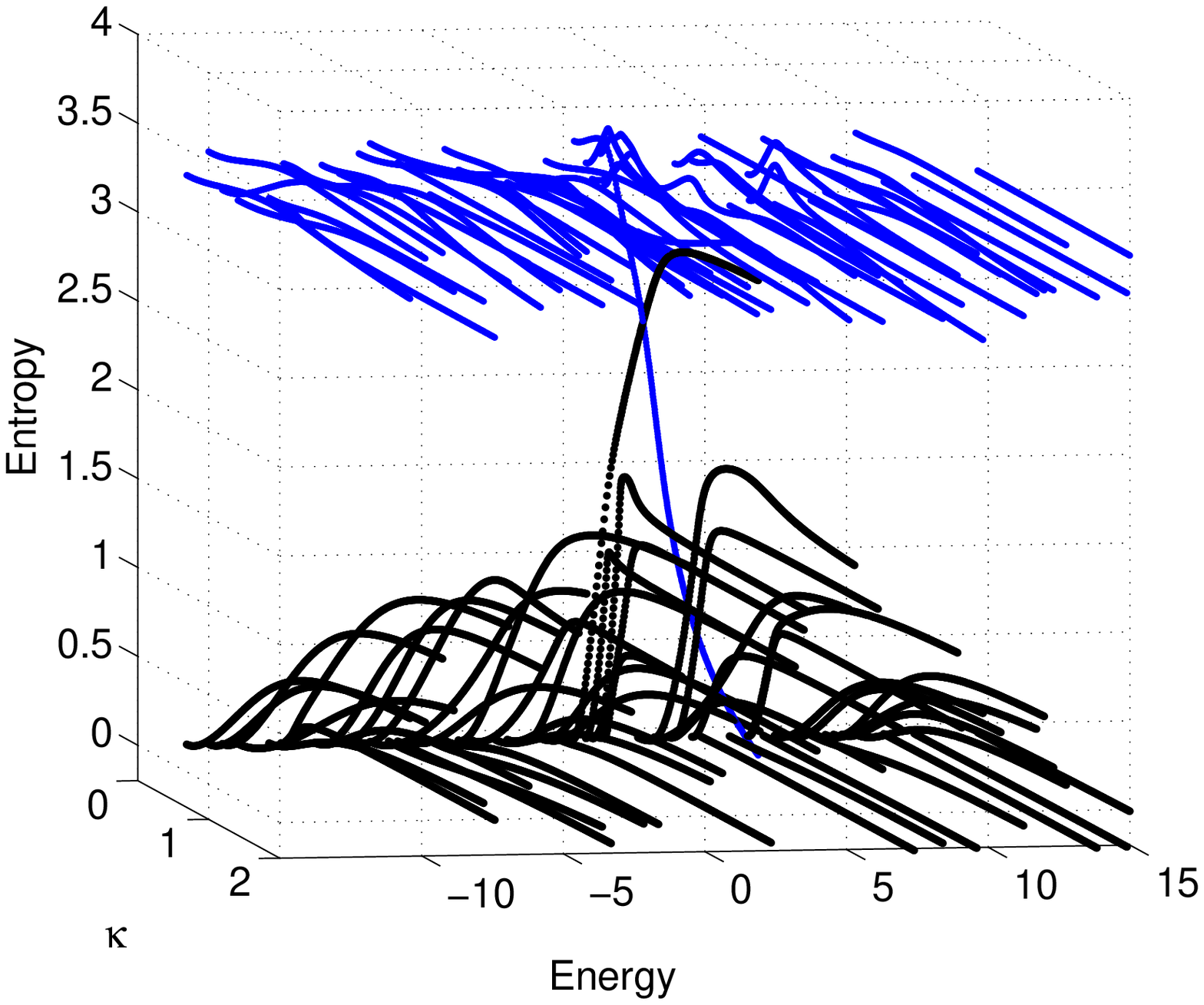}
\caption{\label{fig:entvsk} (Color online) The  evolution of the energy and entropy as $\kappa$ increase for an $N=50$ system.  The blue lines correspond to the basis which the original Hamiltonian $H^0$ is written in. The black lines correspond to the basis in which $H^0$ is diagonal. In both cases, the superradiant state is obvious. Notice that one blue line drops down to low entropy as $\kappa$ increases. One can think of Fig.~\ref{fig:evsk} as a birds eye view of this plot, where we just see the energy and $\kappa$ values. Note that here the GOE  average value of the entropy is $S=3.2$, which is the level most of the blue lines stay at. Conversely the states in the energy basis grow in complexity. Notice how some of the black lines rise to a higher entropy as $\kappa$ increases.  }
\end{figure}

The evolution of the energies with $\kappa$ show robust and interesting features. There are a small number of energies that migrate, then settle down for $\kappa$ in the range  $ 0.5 \rightarrow  1.5$.   In Fig.~\ref{fig:evsk} we see a specific example of this generic behavior. If we look at the entropy of the corresponding wave functions one state in particular emerges.  Starting with a wave function $\psi=\sum _i c_i |n\rangle$ we define the entropy as $S=\sum _i |c_i|^2 \ln(|c_i|^2)$. We can calculate $S$ in the original basis in which we wrote out $H$, or in the energy basis, where $H$ is diagonal and $H^0_{ii}=E_i$. In Fig.~\ref{fig:entvsk} we see the results for an $N=50$ system. The SR state emerges with a very simple structure in the original basis (blue lines), having a very low entropy. The other states stay at the GOE predicted average value $S=\ln (0.48 N)$ \cite{ZeVo2003}, which in this case is 3.2. We see that in the energy basis (black lines), where the Hamiltonian is diagonal for $\kappa=0$, we have the complementary situation with entropy. Now the SR state is a complicated mixture of energy eigenstates with an entropy of around 3, and the other states have lower entropy. Indeed for many states in the energy basis the entropy stays close to zero.

\section{The density of states}\label{sec:dos}
\begin{figure}
\includegraphics[width=.8\textheight]{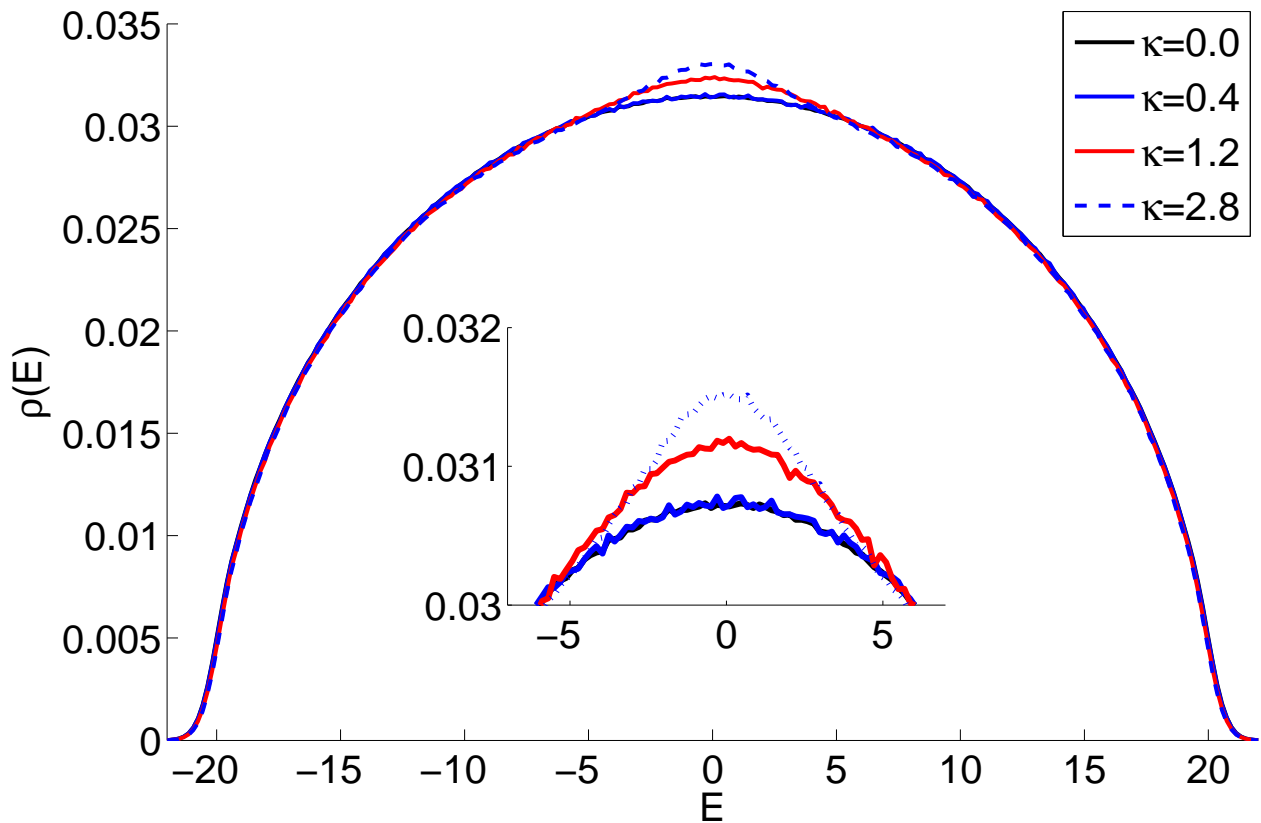}
\caption{\label{fig:rho} (Color online)  Here we  have the density of an ensemble of
2000 matrices with $N=100$ and $\kappa=0,0.4,1.2$ and 2.8.  The level density is close to the semicircle of the GOE even for $\kappa=0.4$. The deviations are consistent with Fig.~\ref{fig:evsk} where levels migrate to the center of the energy range.}
\end{figure}

The migration of a few energies to the middle of the spectrum as $\kappa$ increases is reflected in the density of states. In Fig.~\ref{fig:rho} we see in the  empirical DOS a clear deviation from the semicircle for small $E$. Note we use level density and density of states interchangeably here as there is no degeneracy in the eigenvalues of random matrices.

This change in $\rho(E)$ raises an important practical question for how to do an RMT analysis, mainly how do we unfold the spectrum. First a comment on unfolding spectra. The semicircle level density of the GOE bares no relation to the exponential level density of realistic nuclear systems. To remove the system specific (secular) features of the level density we need to rescale the energies so that the level density is unity across the whole energy range. This process is called unfolding \cite{guhr,brody}. Note that all the fluctuations are preserved even though the unfolded spectrum has a level density of unity. To go from a set of energies $\{E\}$, with density $\rho(E)$ to an unfolded spectrum $\{\xi\}$ with
density $\rho(\xi)= 1$, we need to integrate $\rho(E)$ to get a smooth cumulative level number ${\mathcal N}(E)$:
\begin{equation}
{\mathcal N}(E)=  \int_{-\infty}^{E}\rho(E') dE'.\nonumber\\
\end{equation}
The $i^{th}$ unfolded energy  is simply $\xi_i = {\mathcal N}(E_i)$. In this analysis we unfolded the spectra using the semicircular level density for our convenience.  The results were the same as when we used a numerical fit for the level density.

\section{The width of the energies.}
\label{sec:width}
\begin{figure}
\includegraphics[width=.6\textheight] {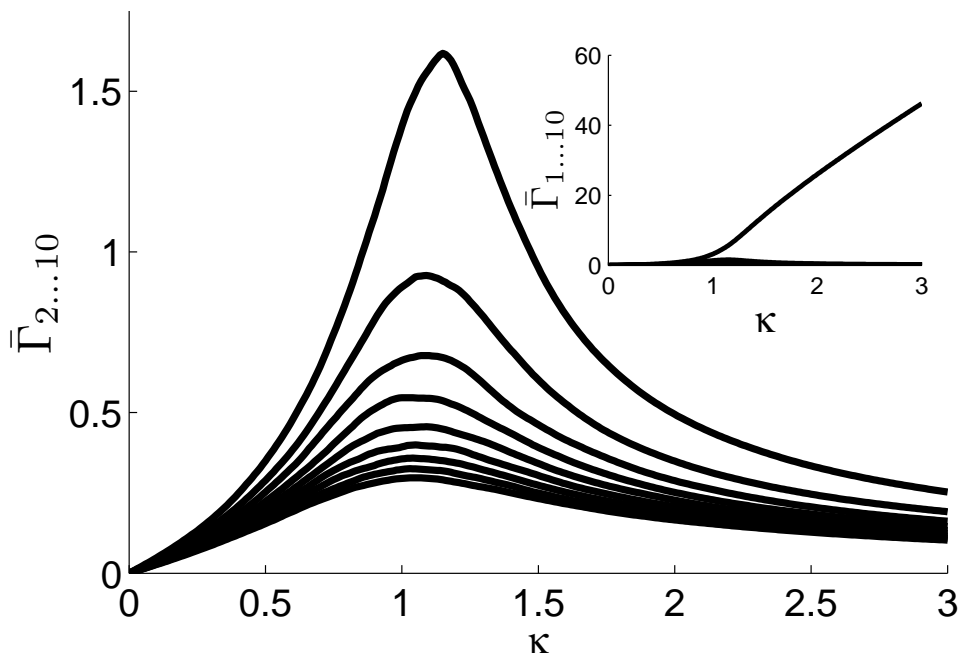}
\caption{\label{fig:gvsk}When $\kappa$  is turned on, the levels
acquire a width, $\Gamma$. Here we sorted the set of 300 $\Gamma$
for each spectrum in an ensemble of 200. For example, $\Gamma_3$ is
the 3rd largest width. The 9 lines in this plot are the ensemble
average of $\Gamma_i$ vs $\kappa$, with $i=2\dots 10$. The insert includes the largest width, $\Gamma_1$ which eventually becomes linear in $\kappa$.}
\end{figure}

\begin{figure}
\includegraphics[width=.6\textheight]{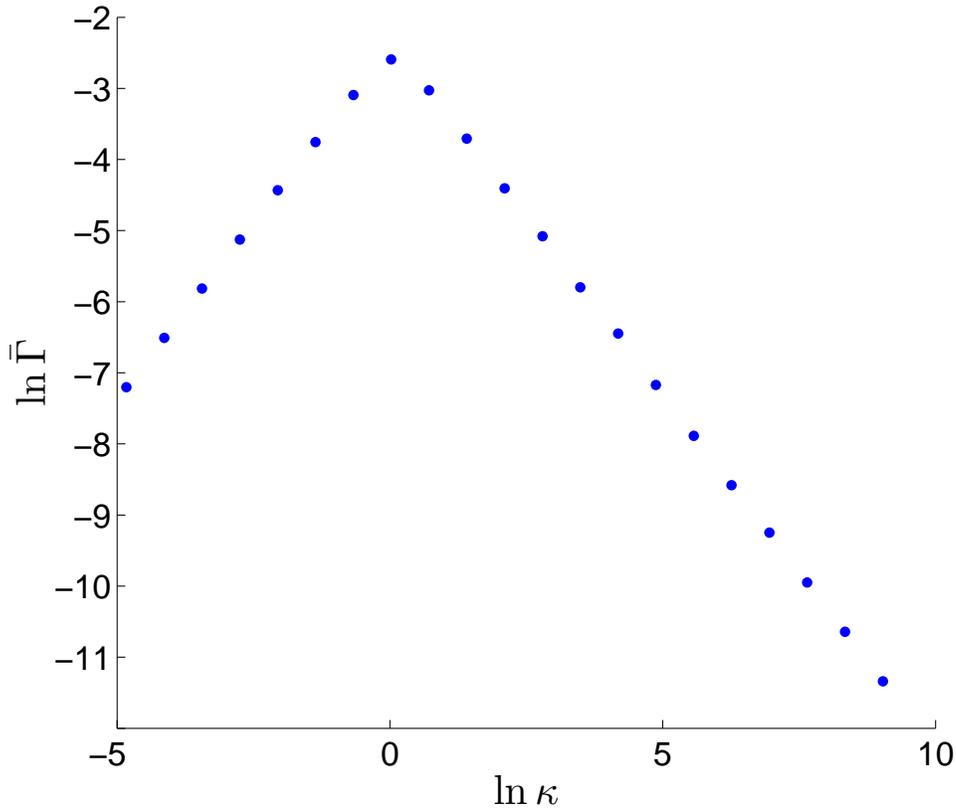}
\caption{\label{fig:lglggam} (Color online) This is a log-log plot of the average of the lines in
Fig.~\ref{fig:gvsk}.}
\end{figure}

The addition of an imaginary part  to $H_{11}$ gives a width to all the levels. These widths can be treated as random variables, and their distribution examined.  An ensemble of 200 matrices was prepared and opened as in Sect.~\ref{sec:opening}. Each random matrix was the start of a sequence of 301 opened matrices with $0 \leq \kappa \leq 3.0$ in steps of 0.01. The complex energies $\varepsilon_n(\kappa)$ were calculated. The widths  $\Gamma(\kappa)$ of the levels  were sorted and their size as a function of $\kappa$ was examined, see  Fig.~\ref{fig:gvsk}. Immediately we see the emergence of the SR state that absorbs all the width. If this state is excluded from the plot we get a completely different behaviour, with the remaining widths having  an exponential dependance on $\kappa$. If we plot $\bar{\Gamma}$, the average of all but the biggest widths, vs $\kappa$ on a log-log plot we get a very simple picture  shown in Fig.~\ref{fig:lglggam}. The straight line sections are roughly $\ln \bar{\Gamma}=\pm \ln \kappa -2.5$. Here the range of $\kappa$ was from $10^{-3}$ to $10^6$. A qualitatively identical SR transition is seen in a different context in \cite{ceka12} where the interplay of disorder and SR was examined in the context of the Anderson model.

\begin{figure}
\includegraphics[width=.8\textheight]{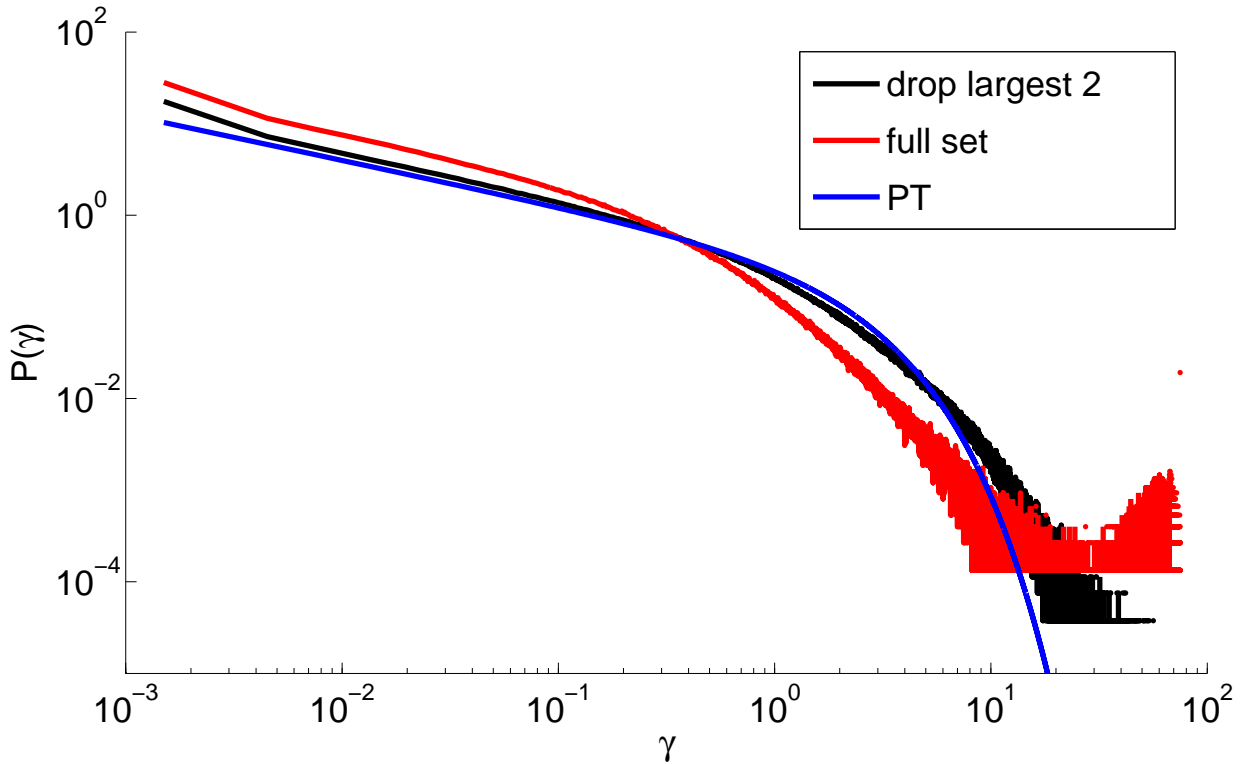} \caption{\label{fig:pgamma}(Color online) The ensemble  result for the distribution of reduced widths $p(\gamma / \bar{\gamma})$.  In red we see the full set of 2.5 million widths which are a  superposition of 250000 matrices with $N=100$ and $\kappa=1.5$. The mean of this set is $\bar{\gamma} = 0.15$. In black we see the results for the  smallest 100 widths of 88646 matrices with $N=102$. The mean of this subset is $\bar{\gamma} = 0.0549$. In blue we have a plot of the function  $P(x)=\frac{1}{\sqrt{2 \pi x}} \exp(-\frac{x}{2})$. }
\end{figure}

This picture of the emergence of the SR state is further reinforced by an analysis of the reduced widths. When the special state is excluded from the analysis we recover the Porter-Thomas distribution (PTD).  In Fig.~\ref{fig:pgamma} we see  a  log-log plot for the
distribution of $\gamma / \bar{\gamma}$ with and without the 2 largest widths for an ensemble with $\kappa=1.5$. We stress that this is as deep as we went in analyzing the distribution of widths. There could well be deviations from the PTD and for a derivation of alternative width distribution see \cite{ShZe12}.

\section{The spectral rigidity, $\Delta_3(L)$.}
\label{sec:d3}
\begin{figure}
\includegraphics[width=0.6\textheight]{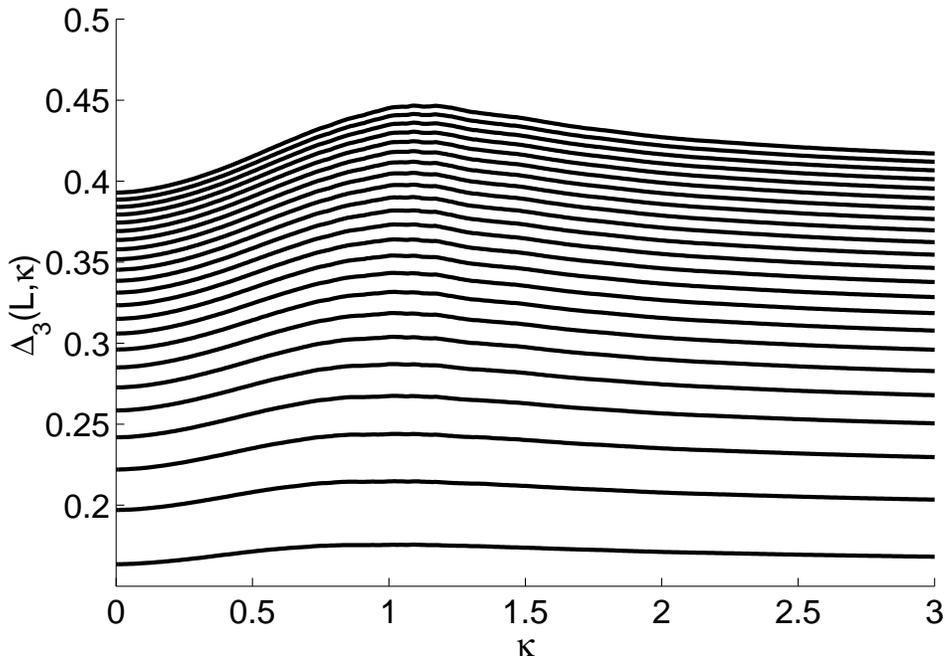} \caption{\label{fig:d3vskens}Here we see  ensemble average of $\Delta_3(L)$ vs $\kappa$. The 23 lines are for the 23 values of $L$, with $5 \leq L \leq 50$ in steps of 2. There are 200 spectra in the ensemble, with $N=300$.}
\end{figure}

\begin{figure}
\includegraphics[width=0.6\textheight]{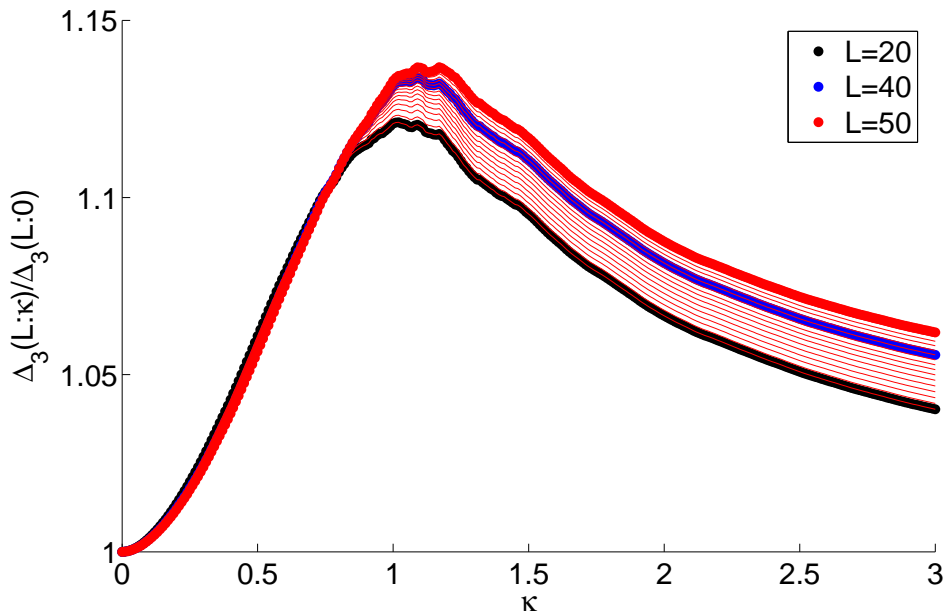}
\caption{\label{fig:d3vskensnrm} (Color online) Here we see  ensemble average of
$\Delta_3(L,\kappa)$ vs $\kappa$ for $20 \leq L \leq 50$, but now
each line is divided by $\Delta_3(L,0)$. The lower $L$ values  are not as
sensitive to $\kappa$ as the window of $L$ levels is too narrow to
probe long range correlations. So it is best to not include the
range $L<20$. The legend refers to the 3 bold lines in the plot. The lines for various $L$ become closer as $L$ increases.}
\end{figure}

The spectral rigidity or  $\Delta_3(L)$ statistic is a common diagnostic for statistical analysis of data based on RMT. It is a robust statistic and can be used to gauge the purity of a spectrum, giving an estimate of the fraction of missed or spurious levels. It can also be used to gauge the degree to which the system is chaotic \cite{shriner07,dyson,brody,me1}. $\Delta_3(L)$ is defined in terms of fluctuations in the cumulative level number, ${\mathcal N}(E)$, the number of levels with energy $\leq E$. ${\mathcal N}(E)$ is a staircase with each step being one unit high, and its slope is the level density $\rho (E)$. A harmonic oscillator will have a regular staircase, with each step being one unit wide. On the other hand the quantum equivalent of a classically regular system has a random but uncorrelated spectrum. In that case ${\mathcal N}(E)$ will have steps whose width have a Poissonian distribution. $\Delta_3(L)$ is a measure of the spectral average deviation of ${\mathcal N}(E)$ from a regular (constant slope) staircase, within an energy interval of length $L$. The spectral average means that the deviation is averaged over the location in the spectrum of the window. The definition is
\begin{eqnarray}
\nonumber \Delta_{3}(L)  &=&\left \langle {\rm min}_{A,B}\; \frac{1}{L}\;\int^{E_i+L}_{E_i}dE'\,[\;{\mathcal N}(E')-AE'-B]^{2}\; \;\right\rangle \\
&=&\langle \delta^i_3(L)  \rangle.\label{eq:d3}
\end{eqnarray}
$A$ and $B$ are calculated  for each $i$ to minimize $\delta_3^i(L)$. The details of the exact calculation of $A$ and $B$ in terms of the energies $\{E_i\}$ are in \cite{me1}. The harmonic oscillator has $\Delta_3(L)=1/12$. At the other extreme, a classically regular system will lead to a quantum mechanical spectrum with no level repulsion. The fluctuations will be far greater because there is no long range correlation giving $\Delta_3(L)=L/15$. The angle brackets mean the average is to be taken over all starting positions $E_i$ of the window of length $L$. This is a spectral average. It is  an amazing fact of RMT that the spectra of the GOE have huge long range correlations, indeed the GOE result is:
\begin{eqnarray}
\Delta_3(L)&=&\frac{1}{\pi^2}\,\left[\log(2\pi L)+\gamma-\frac{5}{4}-\frac{\pi^2}{8}\right]\\
 &=&(\log L-0.0678)/\pi^2
\label{eq:d3th},
\end{eqnarray}
with  $\gamma$ being Euler's constant. We stress that this is the RMT value for the {\sl ensemble average} of $\Delta_3(L)$. The graph of $\Delta_3(L)$ will vary from matrix to matrix but average of many such lines (the ensemble average) will rapidly converge onto Eq. \ref{eq:d3th}. In our opened ensemble, $\Delta_3(L)$ deviates from the GOE result. In  Fig.~\ref{fig:d3vskens} we take various fixed values of $L$ and  we see how the value of $\Delta_3(L)$ evolves with $\kappa$. Deviations from the GOE result ($\kappa=0$) increases slowly as $\kappa$ changes from 0 to 1, then the deviations start to decrease. The effect is similar for all $L$ in the range we looked at, so in Fig.~\ref{fig:d3vskensnrm} we plot all the lines of Fig.~\ref{fig:d3vskens} but divide them by their value at $\kappa=0$. Now we see that $\Delta_3(L)$ increases with $\kappa$ by a similar factor for a broad range of $L$, and furthermore this factor has a maximum value of around 1.15 which happens when $\kappa=1$.

\section{RMT tests for missed levels}
\label{sec:rmt}
The increase in the value of $\Delta_3(L)$ due to opening 0the system could be misconstrued as evidence of spurious or missed levels. There are RMT tests for missed levels and we will apply these tests to the open spectra and see if there is any consistent picture that emerges. We will concentrate on 3 RMT tests, two of them based on $\Delta_3(L)$ and another based on the the nearest neighbor distribution (nnd).

In \cite{mulhall11} a maximum likelihood method based on $\Delta_3(L)$ was developed. The $\Delta_3(L)$ statistic is the spectral average of the set of random numbers $\delta_3^i(L)$. The distribution of these numbers $p(\delta)$ was used as the basis of a likelihood function. The basic idea is that  ${\mathcal N}(\delta)$, the cdf for $p(\delta)$, is a simple function of $\log \delta$.  This led to the following parameterization:
\begin{equation}\label{eq:cdf}
{\mathcal N}(\delta) = \frac{1}{2}(1 - \textrm{Erf}[a + b \log \delta + c
(\log \delta)^2]).
\end{equation}
 An ensemble of depleted spectra was made and the parameters for ${\mathcal N}(\delta)$ were empirically found for a range of $L$, and $x$ the fraction of levels missed, and fitted to smooth functions $a_L(x) ,b_L(x)$ and $c_L(x)$.  Differentiation of ${\mathcal N}(\delta)$ gives probability density for $\delta_3^i(L)$ with $x$ as a continuous parameter:
\begin{equation}\label{eq:plx}
p(\delta,x)=-\frac{1}{\sqrt{\pi }} \exp{[-\big(a_L(x)+b_L(x)\,\log \delta+c_L(x)
\log \delta^2\big) }^2]\,\big(\frac{b_L(x)}{\delta}+\frac{2\ c_L(x) \log \delta}{\delta}\big).
\end{equation}
This is then used as the basis for a maximum likelihood method for determining $x$.

In \cite{bohigas2004}  Bohigas and Pato gave an  expression is given for $\Delta_3(L)$ for incomplete spectra. The fraction of missed levels $x$ is both a scaling factor and a weighting factor and $\Delta_3(L,x)$ is the sum of the GOE and Poissonian result:
\begin{equation}\label{eq:bohigas}
    \Delta_3(L,x)=x^2\frac{L/x}{15}+(1-x)^2\Delta_3^{\textrm{GOE}}(L/(1-x)).
\end{equation}
The $\Delta_3(L)$ statistic of an open spectrum can be compared with this expression and the best $x$ found.

The nearest neighbor distribution (nnd) is another commonly used statistic. The nnd for a pure spectrum follows the Wigner distribution,
\begin{equation}
    P(s)=\frac{\pi}{2}se^{-\pi s^2/4},
\end{equation}
where $s=S/D$, $S$ being the spacing between adjacent levels, and $D$ is the average spacing ($D=1$ for an unfolded spectrum).  The nnd of a spectrum incomplete by a fraction $x$ is given by
\begin{equation}
P(s)=\sum_{k=0}^{\infty} (1-x) x^k P(k;s);\label{eq:pofsx}
\end{equation}
where $P(k;s)$ is the $k^{th}$ nearest neighbor spacing, $E_{k+i}-E_i$. This was first introduced as an ansatz in \cite{watson81}, and rederived in  \cite{agv} and   \cite{bohigas2004}.  Eq.~\ref{eq:pofsx} was used by Agvaanluvsan et al  as the basis for a maximum likelihood method (MLM) to determine $x$ for incomplete spectra \cite{agv}.

\begin{figure}
\includegraphics[width=.6\textheight]{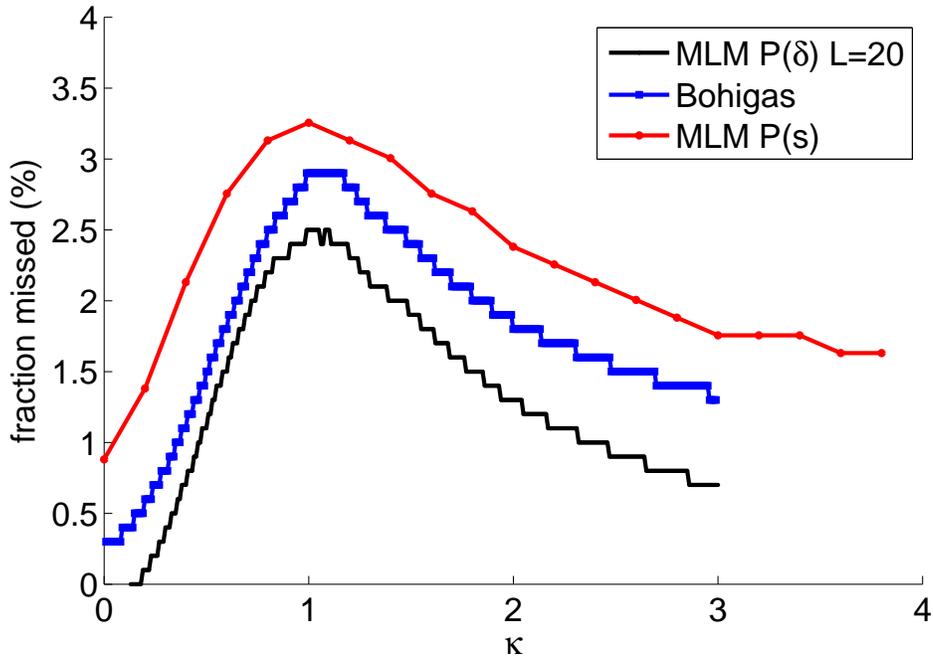}
\caption{\label{fig:mlm}(Color online) Here we see the results of tests to determine the fraction of levels missed in incomplete spectra vs $\kappa$ for open but complete GOE spectra. $N$ is 300 in all cases.}
\end{figure}

These three tests for missed levels were applied to complete opened GOE spectra of dimension $N=300$. The value of $\kappa$ went from 0 to 3 in steps of 0.01. The values $x$ for the fraction depleted vs $\kappa$  are shown in Fig.~\ref{fig:mlm}. It appears that the spectra look incomplete when the system is opened, and the effect is strongest  to the tune of about 3\% when $\kappa=1$.

\section{Conclusion}
\label{sec:conc}
 A simple model for an open quantum system was realized by adding an imaginary number $\imath \sqrt{N} \kappa$ to the trace of an $N\times N$  GOE matrix. The level density deviated from the Wigner semicircle  and the deviation grew with $\kappa$. There was a drifting of some levels to the center of the spectrum at around $\kappa=1$. A very low entropy state emerged which  we identifed with a super-radiant state.   The widths  $\Gamma(\kappa)$ of the levels  were sorted and a graph of the biggest 10 showed the emergence of this SR state that absorbs all the width. When the largest width is excluded the remaining widths were consistent with a Porter-Thomas distribution. Their average value had a simple exponential dependance on $\kappa$.  A plot of $\bar{\Gamma}$, the average of all but the biggest widths vs $\kappa$ on a log-log plot reveals a SR transition. The $\Delta_3(L)$ statistic deviated from the GOE value and looked like that of an incomplete spectra, or an impure spectra with intruder levels. Three separate tests for missed levels based on RMT consistently showed a that at $\kappa \approx 1$ the spectra appeared incomplete or contaminated to the tune of about 3\%.

It is interesting that such a simple system can capture so much of the systematics of super-radiance in open quantum systems.

\begin{acknowledgments}
We wish to acknowledge the support of the Office of Research Services of the University of Scranton, Vladimir Zelevinsky and Alexander Volya for the original idea and for valuable discussions throughout. Yan V Fyodorov made useful comments regarding  the work done on the distribution of complex energies in chaotic quantum systems. The anonymous referee significantly improved the clarity of the presentation with constructive comments.

\end{acknowledgments}

\bibliography{OpenSys}

\begin{thebibliography}{19}
\expandafter\ifx\csname natexlab\endcsname\relax\def\natexlab#1{#1}\fi
\expandafter\ifx\csname bibnamefont\endcsname\relax
  \def\bibnamefont#1{#1}\fi
\expandafter\ifx\csname bibfnamefont\endcsname\relax
  \def\bibfnamefont#1{#1}\fi
\expandafter\ifx\csname citenamefont\endcsname\relax
  \def\citenamefont#1{#1}\fi
\expandafter\ifx\csname url\endcsname\relax
  \def\url#1{\texttt{#1}}\fi
\expandafter\ifx\csname urlprefix\endcsname\relax\def\urlprefix{URL }\fi
\providecommand{\bibinfo}[2]{#2}
\providecommand{\eprint}[2][]{\url{#2}}

\bibitem[{\citenamefont{Sokolov and Zelevinsky}(1988)}]{soze88}
\bibinfo{author}{\bibfnamefont{V.}~\bibnamefont{Sokolov}} \bibnamefont{and}
  \bibinfo{author}{\bibfnamefont{V.}~\bibnamefont{Zelevinsky}},
  \bibinfo{journal}{Physics Letters B} \textbf{\bibinfo{volume}{202}},
  \bibinfo{pages}{10} (\bibinfo{year}{1988}).

\bibitem[{\citenamefont{Sokolov and Zelevinsky}(1992)}]{soze92}
\bibinfo{author}{\bibfnamefont{V.}~\bibnamefont{Sokolov}} \bibnamefont{and}
  \bibinfo{author}{\bibfnamefont{V.}~\bibnamefont{Zelevinsky}},
  \bibinfo{journal}{Annals of Physics} \textbf{\bibinfo{volume}{216}},
  \bibinfo{pages}{323 } (\bibinfo{year}{1992}), ISSN \bibinfo{issn}{0003-4916},
  \urlprefix\url{http://www.sciencedirect.com/science/article/pii/000349169290180T}.

\bibitem[{\citenamefont{Zelevinsky and Volya}(2004)}]{ZeVo2003}
\bibinfo{author}{\bibfnamefont{V.}~\bibnamefont{Zelevinsky}} \bibnamefont{and}
  \bibinfo{author}{\bibfnamefont{A.}~\bibnamefont{Volya}},
  \bibinfo{journal}{Physics Reports} \textbf{\bibinfo{volume}{391}},
  \bibinfo{pages}{311 } (\bibinfo{year}{2004}), ISSN \bibinfo{issn}{0370-1573},
  \bibinfo{note}{from atoms to nuclei to quarks and gluons: the omnipresent
  manybody theory},
  \urlprefix\url{http://www.sciencedirect.com/science/article/pii/S0370157303004319}.

\bibitem[{\citenamefont{Mulhall}(2011)}]{mulhall11}
\bibinfo{author}{\bibfnamefont{D.}~\bibnamefont{Mulhall}},
  \bibinfo{journal}{Phys. Rev. C} \textbf{\bibinfo{volume}{83}},
  \bibinfo{pages}{054321} (\bibinfo{year}{2011}),
  \urlprefix\url{http://link.aps.org/doi/10.1103/PhysRevC.83.054321}.

\bibitem[{\citenamefont{Shriner et~al.}(2007)\citenamefont{Shriner, Pato,
  Mitchell, and Tufaile}}]{shriner07}
\bibinfo{author}{\bibfnamefont{J.~F.} \bibnamefont{Shriner}},
  \bibinfo{author}{\bibfnamefont{M.}~\bibnamefont{Pato}},
  \bibinfo{author}{\bibfnamefont{G.}~\bibnamefont{Mitchell}}, \bibnamefont{and}
  \bibinfo{author}{\bibfnamefont{A.}~\bibnamefont{Tufaile}},
  \bibinfo{journal}{Nuclear Instruments and Methods in Physics Research Section
  A: Accelerators, Spectrometers, Detectors and Associated Equipment}
  \textbf{\bibinfo{volume}{581}}, \bibinfo{pages}{831 } (\bibinfo{year}{2007}),
  ISSN \bibinfo{issn}{0168-9002},
  \urlprefix\url{http://www.sciencedirect.com/science/article/pii/S0168900207018736}.

\bibitem[{\citenamefont{Sommers et~al.}(1999)\citenamefont{Sommers, Fyodorov,
  and Titov}}]{fyodorov99a}
\bibinfo{author}{\bibfnamefont{H.-J.} \bibnamefont{Sommers}},
  \bibinfo{author}{\bibfnamefont{Y.~V.} \bibnamefont{Fyodorov}},
  \bibnamefont{and} \bibinfo{author}{\bibfnamefont{M.}~\bibnamefont{Titov}},
  \bibinfo{journal}{Journal of Physics A: Mathematical and General}
  \textbf{\bibinfo{volume}{32}}, \bibinfo{pages}{L77} (\bibinfo{year}{1999}),
  \urlprefix\url{http://stacks.iop.org/0305-4470/32/i=5/a=003}.

\bibitem[{\citenamefont{Fyodorov and Sommers}(1996)}]{fyodorov96}
\bibinfo{author}{\bibfnamefont{Y.}~\bibnamefont{Fyodorov}} \bibnamefont{and}
  \bibinfo{author}{\bibfnamefont{H.-J.} \bibnamefont{Sommers}},
  \bibinfo{journal}{Journal of Experimental and Theoretical Physics Letters}
  \textbf{\bibinfo{volume}{63}}, \bibinfo{pages}{1026} (\bibinfo{year}{1996}),
  ISSN \bibinfo{issn}{0021-3640},
  \urlprefix\url{http://dx.doi.org/10.1134/1.567120}.

\bibitem[{\citenamefont{Fyodorov and Khoruzhenko}(1999)}]{fyodorov99b}
\bibinfo{author}{\bibfnamefont{Y.~V.} \bibnamefont{Fyodorov}} \bibnamefont{and}
  \bibinfo{author}{\bibfnamefont{B.~A.} \bibnamefont{Khoruzhenko}},
  \bibinfo{journal}{Phys. Rev. Lett.} \textbf{\bibinfo{volume}{83}},
  \bibinfo{pages}{65} (\bibinfo{year}{1999}),
  \urlprefix\url{http://link.aps.org/doi/10.1103/PhysRevLett.83.65}.

\bibitem[{\citenamefont{Auerbach and Zelevinsky}(2011)}]{AuZe2011}
\bibinfo{author}{\bibfnamefont{N.}~\bibnamefont{Auerbach}} \bibnamefont{and}
  \bibinfo{author}{\bibfnamefont{V.}~\bibnamefont{Zelevinsky}},
  \bibinfo{journal}{Rep. Prog. Phys.} \textbf{\bibinfo{volume}{74}}
  (\bibinfo{year}{2011}).

\bibitem[{\citenamefont{Celardo and Kaplan}(2009)}]{ceka09}
\bibinfo{author}{\bibfnamefont{G.~L.} \bibnamefont{Celardo}} \bibnamefont{and}
  \bibinfo{author}{\bibfnamefont{L.}~\bibnamefont{Kaplan}},
  \bibinfo{journal}{Phys. Rev. B} \textbf{\bibinfo{volume}{79}},
  \bibinfo{pages}{155108} (\bibinfo{year}{2009}),
  \urlprefix\url{http://link.aps.org/doi/10.1103/PhysRevB.79.155108}.

\bibitem[{\citenamefont{Guhr et~al.}(1998)\citenamefont{Guhr, Mueller-Groeling,
  and Weidenmueller}}]{guhr}
\bibinfo{author}{\bibfnamefont{T.}~\bibnamefont{Guhr}},
  \bibinfo{author}{\bibfnamefont{A.}~\bibnamefont{Mueller-Groeling}},
  \bibnamefont{and} \bibinfo{author}{\bibfnamefont{H.~A.}
  \bibnamefont{Weidenmueller}}, \bibinfo{journal}{Phys.\ Rep.}
  \textbf{\bibinfo{volume}{299}}, \bibinfo{pages}{189} (\bibinfo{year}{1998}).

\bibitem[{\citenamefont{Brody et~al.}(1981)\citenamefont{Brody, Flores, French,
  Mello, Pandey, and Wong}}]{brody}
\bibinfo{author}{\bibfnamefont{T.~A.} \bibnamefont{Brody}},
  \bibinfo{author}{\bibfnamefont{J.}~\bibnamefont{Flores}},
  \bibinfo{author}{\bibfnamefont{J.~B.} \bibnamefont{French}},
  \bibinfo{author}{\bibfnamefont{P.~A.} \bibnamefont{Mello}},
  \bibinfo{author}{\bibfnamefont{A.}~\bibnamefont{Pandey}}, \bibnamefont{and}
  \bibinfo{author}{\bibfnamefont{S.~S.~M.} \bibnamefont{Wong}},
  \bibinfo{journal}{Rev. Mod. Phys.} \textbf{\bibinfo{volume}{53}},
  \bibinfo{pages}{385} (\bibinfo{year}{1981}).

\bibitem[{\citenamefont{Celardo et~al.}(2013)\citenamefont{Celardo, Biella,
  Kaplan, and Borgonovi}}]{ceka12}
\bibinfo{author}{\bibfnamefont{G.}~\bibnamefont{Celardo}},
  \bibinfo{author}{\bibfnamefont{A.}~\bibnamefont{Biella}},
  \bibinfo{author}{\bibfnamefont{L.}~\bibnamefont{Kaplan}}, \bibnamefont{and}
  \bibinfo{author}{\bibfnamefont{F.}~\bibnamefont{Borgonovi}},
  \bibinfo{journal}{Fortschritte der Physik} \textbf{\bibinfo{volume}{61}},
  \bibinfo{pages}{250} (\bibinfo{year}{2013}), ISSN \bibinfo{issn}{1521-3978},
  \urlprefix\url{http://dx.doi.org/10.1002/prop.201200082}.

\bibitem[{\citenamefont{Shchedrin and Zelevinsky}(2012)}]{ShZe12}
\bibinfo{author}{\bibfnamefont{G.}~\bibnamefont{Shchedrin}} \bibnamefont{and}
  \bibinfo{author}{\bibfnamefont{V.}~\bibnamefont{Zelevinsky}},
  \bibinfo{journal}{Phys. Rev. C} \textbf{\bibinfo{volume}{86}},
  \bibinfo{pages}{044602} (\bibinfo{year}{2012}),
  \urlprefix\url{http://link.aps.org/doi/10.1103/PhysRevC.86.044602}.

\bibitem[{\citenamefont{Dyson and Mehta}(1963)}]{dyson}
\bibinfo{author}{\bibfnamefont{F.~J.} \bibnamefont{Dyson}} \bibnamefont{and}
  \bibinfo{author}{\bibfnamefont{M.~L.} \bibnamefont{Mehta}},
  \bibinfo{journal}{J.\ Math.\ Phys} \textbf{\bibinfo{volume}{4}},
  \bibinfo{pages}{701} (\bibinfo{year}{1963}).

\bibitem[{\citenamefont{Mulhall et~al.}(2000)\citenamefont{Mulhall, Volya, and
  Zelevinsky}}]{me1}
\bibinfo{author}{\bibfnamefont{D.}~\bibnamefont{Mulhall}},
  \bibinfo{author}{\bibfnamefont{A.}~\bibnamefont{Volya}}, \bibnamefont{and}
  \bibinfo{author}{\bibfnamefont{V.}~\bibnamefont{Zelevinsky}},
  \bibinfo{journal}{Phys.\ Rev.\ Lett.} \textbf{\bibinfo{volume}{85}}
  (\bibinfo{year}{2000}).

\bibitem[{\citenamefont{Bohigas and Pato}(2004)}]{bohigas2004}
\bibinfo{author}{\bibfnamefont{O.}~\bibnamefont{Bohigas}} \bibnamefont{and}
  \bibinfo{author}{\bibfnamefont{M.~P.} \bibnamefont{Pato}},
  \bibinfo{journal}{Physics Letters B} \textbf{\bibinfo{volume}{595}},
  \bibinfo{pages}{171 } (\bibinfo{year}{2004}), ISSN \bibinfo{issn}{0370-2693},
  \urlprefix\url{http://www.sciencedirect.com/science/article/B6TVN-4CVX10M-3/2/340546fa17309004a02087ec474d7cfa}.

\bibitem[{\citenamefont{Watson~III et~al.}(1981)\citenamefont{Watson~III,
  Bilpuch, and Mitchell}}]{watson81}
\bibinfo{author}{\bibfnamefont{W.~A.} \bibnamefont{Watson~III}},
  \bibinfo{author}{\bibfnamefont{E.~G.} \bibnamefont{Bilpuch}},
  \bibnamefont{and} \bibinfo{author}{\bibfnamefont{G.~E.}
  \bibnamefont{Mitchell}}, \bibinfo{journal}{Z. Phys.}
  \textbf{\bibinfo{volume}{A300}}, \bibinfo{pages}{89} (\bibinfo{year}{1981}).

\bibitem[{\citenamefont{Agvaanluvsan et~al.}(2003)\citenamefont{Agvaanluvsan,
  Mitchell, Shriner~Jr., and Pato}}]{agv}
\bibinfo{author}{\bibfnamefont{U.}~\bibnamefont{Agvaanluvsan}},
  \bibinfo{author}{\bibfnamefont{G.~E.} \bibnamefont{Mitchell}},
  \bibinfo{author}{\bibfnamefont{J.~F.} \bibnamefont{Shriner~Jr.}},
  \bibnamefont{and} \bibinfo{author}{\bibfnamefont{M.~P.} \bibnamefont{Pato}},
  \bibinfo{journal}{NIMA} \textbf{\bibinfo{volume}{498}}, \bibinfo{pages}{459}
  (\bibinfo{year}{2003}).

\end{thebibliography}

\end{document}